\documentstyle[12pt,aps]{revtex}

\begin{document}

\title
{Orbital Magnetism in Small Quantum Dots with Closed Shells}
\author
{W.D. Heiss$^{\star }$ and R.G. Nazmitdinov$ ^{\star \dagger}$}
\address
{$^{\star }$Centre for Nonlinear Studies and Department of Physics, \\
University of the Witwatersrand, PO Wits 2050, Johannesburg, South Africa \\
$^{\dagger}$ Joint Institute for Nuclear Research, 141980 Dubna, 
Russia}
\maketitle

Pacs: 73.20Dx, 73.23Ps

\begin{abstract}
It is found that various kind of shell structure which occurs at
specific values of the magnetic field leads to the disappearance of the
orbital magnetization for particular magic numbers of
small quantum dots with an electron number $A < 30$.
\end{abstract}

The development of semiconductor technology has made possible
the confinement of a finite number of electrons in a localized
space of a few hundred Angstroms \cite{Kas}.
These mesoscopic systems called quantum dots
open new avenues in the study of the interplay between
quantum and classical behavior at a low--dimensional scale.
The smaller the quantum dot, the larger
the prevalence of quantum effects upon the static and dynamic properties
of the system. Their electronic properties are determined by the
interplay of the external confinement and the electron-electron interaction
which produces the effective mean field of the "artificial atom"
\cite{MC,rev1}.

The quasiparticle concept associated with an effective mean field is well
established in many particle physics. For finite Fermi system
like nuclei or metallic clusters the bunching of single particle levels known
as shells \cite{NR,HB} is one consequence of this description, if the
mean free path of the particles is comparable with the size of the system.
A remarkable stability is found
in nuclei and metallic clusters at magic numbers which correspond to closed
shells in the effective potential. For small quantum dots, where
the number of electrons is well defined $(A<30)$, the mean free path
of the electrons appears to be comparable with the diameter of the dot.
Transport phenomena are governed by the physics of the
Coulomb blockade regime \cite{rev1}. In recent experiments \cite{e1,e2,e3}
shell structure effects have been observed clearly for quantum dots.
In particular, the energy needed to place the extra electron (addition energy)
into a vertical quantum dot at zero magnetic field
has characteristic maxima which correspond to
the sequence of magic numbers of a two-dimensional harmonic
oscillator \cite{e2}. The energy gap between
filled shells is $\hbar \omega_0$,
where $\hbar \omega_0$ is the lateral confinement energy.
In fact, when the confining energy is comparable to or larger than the
interaction energy, these atomic-like features  have been predicted in a
number of publications \cite{Mac}-\cite{Sto}. While the electron-electron
interaction is important for the explanation of certain
ground state properties like special values of angular momenta
of quantum dots in a magnetic field \cite{MC}, for small
number of electrons the confinement energy becomes prevalent over the Coulomb
energy \cite{Mac,Sto,DN}. In \cite{W} it was demonstrated
that the magnetoexciton spectrum in small quantum dots
resembles well the spectrum of the noninteracting electron--hole
pairs. In particular, the gaps in the spectrum, which are
typical features of the shell structure, reappear at different
values of the magnetic field. 
Recent calculations using the spin--density-functional approach
\cite{ken} nicely confirm shell closure for small magic electron numbers
in a parabolic quantum dot.

Orbital magnetism of an ensemble of quantum dots         
has been discussed for non--interacting electrons in \cite{al,op,ric},
but little attention was paid to shell structure of an individual dot.
We demonstrate within a simple model
that the disappearance and re-appearance of closed shells in a quantum dot
under variation of the magnetic field strength leads to a novel
feature: the orbital magnetization disappears for particular values of
the magnetic field strength, which are associated to particular
magic numbers. 

Since the electron interaction is crucial only
for partially filled electronic shells \cite{ken,W1}, we deal in this paper
mainly with closed shells. It corresponds to the quantum limit
${\hbar \omega_0}\geq e^2/\varepsilon l_0$ where $e^2/\varepsilon l_0$
is the typical Coulomb energy with $ l_0=(\hbar/m^*\omega_0)^{1/2}$,
$m^*$ is the effective electron mass
and $\varepsilon $ is the dielectric constant.
In fact, for small dots, where large gaps
between closed shells occur \cite{Mac,Sto,W}, the electron interaction plays
the role of a weak perturbation which can be neglected.
But even in the regime ${\hbar \omega_0}<  e^2/\varepsilon l_0$ 
a distinctively larger addition energy is needed, if an electron is added  
to a closed shell \cite{ken}.
We choose the harmonic oscillator potential as the effective mean field for
the electrons in an isolated quantum dot.
Our discussion here is based upon the 2D version of the Hamiltonian
\cite{HeNa97} including spin degree of freedom. The magnetic field acts
perpendicular to the plane of motion, i.e.
 $H=\sum_{j=1}^A h_j$
with
\begin{equation}
h={1\over 2m^*}(\vec p-{e\over c}\vec A)^2+{m^*\over 2}(\omega _x^2 x^2
+\omega _y^2 y^2) + \mu^*\sigma _zB.           \label{ham}
\end{equation}
where $\vec A=[\vec r \times \vec B]/2, \, \vec B=(0,0,B)$ and
$\sigma _z$ is the Pauli matrix.
We do not take into account the effect of finite temperature; this is
appropriate for experiments which are performed at temperatures
$kT\ll \Delta$ with $\Delta $ being the mean level spacing. The units used 
are meV for the energy and Tesla for the magnetic field strength. The
effective mass is $m^*=0.067m_e$ for GaAs, which yields, for $A\approx 15$,
the size $R_0\approx 320 \AA$ and $\hbar \omega_0=3meV$ \cite{HeNa97}. 
The effective mass determines the orbital magnetic moment for the electrons 
and leads to  $\mu _B^{{\rm eff}}= m_e/m^* \mu_B \approx 15\mu _B$.
The effective spin magnetic moment is $\mu ^*=g_L\mu _B$ 
 with the effective Land\'e factor $g_L=0.44$
and $\mu_B=|e|\hbar/2m_e c$. The
magnetic orbital effect is much enhanced in comparison with the magnetic spin
effect, yet the tiny spin splitting does produce signatures as we see below.

Shell structure occurs whenever the ratio of the two eigenmodes
$\Omega _{\pm }$ of the Hamiltonian (\ref{ham}) (see Ref.\cite{HeNa97})
\begin{equation}
\label{mod}
\Omega _{\pm }^2={1\over 2}(\omega _x^2+\omega _y^2+4\omega _L^2\pm 
\sqrt{(\omega _x^2-\omega _y^2)^2+8\omega _L^2(\omega _x^2+\omega _y^2)
 + 16\omega _L^4})
\end{equation}
is a rational number with a small numerator and denominator.
Here $\omega_L= |e|B/(2m^*c)$.
Closed shells are particularly pronounced if
the ratio is equal to one (for $B=0$) or two (for
$B\approx 1.23$ ) or three (for $B\approx 2.01$ ) and lesser
pronounced if the ratio is 3/2 (for $B=0.72$) or 5/2 (for $B=1.65$)
for a circular case $\omega_x=\omega_y$ (see Fig.1a).
Note that, for better illustration, we used for the spin splitting
the value $2 \mu _B$ instead of the correct $\mu ^*$ in all Figures; the
discussions and conclusions are based on the correct value.
The values given here for $B$ depend on $m^*$ and $\omega _{x,y}$.
As a consequence, a material with an even smaller effective mass $m^*$ would
show these effects for a correspondingly smaller magnetic field.
For $B=0$ the magic numbers (including spin) turn out to be the
usual sequence of the two dimensional isotropic oscillator, that is
$2,6,12,20,\ldots $ \cite{e2}. For $B\approx 1.23$
we find new shells {\em as if} the confining potential would be a
deformed harmonic oscillator without magnetic field.
The magic numbers are
$2,4,8,12,18,24,\ldots $ which are just the numbers obtained from the two
dimensional oscillator with $\omega _>=2\omega _<$ ($\omega _>$ and
$\omega _<$ denote the larger and smaller value of the two frequencies).
Similarly, we get for $B\approx 2.01$ the magic numbers
$2,4,6,10,14,18,24,\ldots $ which corresponds to $\omega _>=3\omega _<$.
If we start from the outset with a deformed mean field
$\omega _x=(1-\beta )\omega _y$ with $\beta >0$,
the degeneracies (closed shells) lifted at $B=0$
re-occur at higher values for $B$ (see Fig.2 and discussion relating to it).
In Fig.1b we display an example
referring to $\beta =0.2$. The significance of this finding lies in the
restoration of closed shells by the magnetic field in an isolated
quantum dot that
does not give rise to magic numbers at zero field strength due to deformation.
We mention that the choice $\beta =0.5$ would shift the pattern found at
$B\approx 1.23$ in Fig.1a to the value $B=0$. 
The relation between $B$ and
the deformation is displayed in Fig.2, where, for better illustration,
$B^{'}=\omega_L/\omega_x$ rather than $B$ is plotted {\it versus}
$r=\omega _x/\omega _y$.  Closed shells are
obtained for values of $B$ and $\beta $ which yield
$\Omega _+/\Omega _-=k=1,2,3,\ldots $, that is for values on 
the trajectories of Fig.2.

 Note also the asymmetry of
the curves in Fig.2: while $\omega _x/\omega _y=r$ is physically identical
with $\omega _x/\omega _y=1/r$ without magnetic field, the two deformations
become distinct by the presence of a magnetic field as it establishes
a direction perpendicular to the $x-y-$plane.

In \cite{HeNa97} we have obtained various shapes of the
quantum dot by energy minimization. In this context it is worth noting that
at the particular values of the magnetic field, where a closed shell
occurs, the energy minimum would be obtained for circular dots,
if the particle number is chosen to be equal to the magic numbers. Deviations
from those magic numbers usually give rise to deformed shapes at the energy
minimum. To what extent these 'spontaneous' deformations actually occur
(which is the Jahn--Teller effect \cite{JT}), is subject to
more detailed experimental information.
The far-infrared spectroscopy in a small isolated quantum dot
could be a useful tool to provide pertinent data \cite{HeNa97}.

The question arises as to what extent our findings depend upon the particular
choice of the mean field. 
The Coulomb interaction lowers the electron levels for increasing magnetic
quantum number $|m|$ \cite{MC,DN}.
The addition of the term $-\lambda \hbar \omega L^2$ 
to the Hamiltonian (\ref{ham}), 
where $L$ is the dimensionless $z$-component of
the angular momentum operator, mimics
this effect for $\lambda >0$ in the Coulomb blockade regime of
deformed quantum dots \cite{hac}. 
In this way, it interpolates the single-particle spectrum between that of the
oscillator and the square well \cite{NR}.
For $\omega _x\ne \omega _y$ and $\lambda \ne 0$ the 
Hamiltonian $H'=H - \lambda \hbar \omega L^2$
is non-integrable \cite{HeNa94} and the
level crossings encountered in Figs.1 become avoided level crossings. 
The shell structure, which prevails for $\lambda =0$ throughout
the spectrum at $B\approx 1.23$ or $B\approx 2.01$, is therefore disturbed 
to an increasing extent with increasing shell number. But even for 
$\lambda \le 0.1$ the structure is still clearly discernible 
for about seven shells, that is for particle numbers up to about twenty five.

When the magnetic field is changed continuously for a quantum dot of fixed
electron number, the ground state will undergo a rearrangement at the
values of $B$, where level crossings occur \cite{MC,DN}.
In fact, it leads to strong variations in the magnetization \cite{MC} and
should be observable also in
the magnetic susceptibility as it is proportional to the second derivative
of the total energy with respect to the field strength.
While details may be modified by electron correlations, we think
that the general features discussed below should be preserved.

In Fig.3 we discern clearly distinct patterns depending on
the electron number, in fact, the susceptibility appears to be a fingerprint
pertaining to the electron number.
Deforming the oscillator does not produce new features except for the
fact that all lines in Fig.3 would be shifted in
accordance with Fig.2. If there is no level crossing, the
second derivative of $E_{{\rm tot}}$ is a smooth function. The crossing of
two occupied levels does not change the smoothness. In contrast, if an
unoccupied level crosses the last occupied level,
the second derivative of $E_{{\rm tot}}$ must
show a spike.
In this way, we understand the even-odd effect when comparing $A=8$
with $A=9$ in Fig.3. The spin splitting
caused by the magnetic field at $B\approx 2.01$ for $A=8$
is absent for $A=9$. This becomes evident when looking at a blow up of
this particular level crossing which is illustrated in Fig.4, where the last
occupied level is indicated as a thick line and the points where a spike
occurs are indicated by a dot.
Note that the splitting is proportional to the effective spin
magnetic moment $\mu ^*$.

Spikes of the susceptibility are associated with a spin flip
for even electron numbers. They are brought about by
the crossing of the top (bottom) with the bottom (top) line of a
double line. Hence, both lines of the double splitting
in Fig.3  yield a spin flip ($A=8$), but neither of the single
lines ($A=9$). Strictly speaking, the spikes are $\delta$-functions with a
factor which is determined by the angle at which the two relevant lines
cross.    If the level
crossings are replaced by avoided crossings (Landau-Zener crossings), the
lines would be broadened. This would be the case in the present model
for $\lambda >0$ {\em and} $\beta >0$. Finite temperature will
also result in line broadening.

We now focus on the special cases which give rise to closed shells,
that is when the ratio $\Omega_+/\Omega_-=k=1,2,3,\ldots $.
For the sake of clarity we analyze in detail the circular shape
($\omega_x=\omega_y=\omega_0$) for which the eigenmodes (Eq.(\ref{mod}))
become $\Omega_{\pm} = (\Omega \pm \omega_L )$ with
$\Omega = \sqrt{\omega_0^2+\omega_L^2}$ \cite{Fock}. We find for the
magnetization
\begin{equation}
\label{mag1}
M= \mu_B^{{\rm eff}}(1-{\omega_L \over \Omega})
({\sum}_{-}-k {\sum}_{+})- \mu^*<S_z>
\end{equation}
with  $\sum_{\pm } = \sum _j^A(n_{\pm }+1/2)_j$ \cite{HeNa97}.
For completely filled shell $<S_z>=0$, since, for the magnetic field
strengths considered here, the spin orientations cancel each other
(see Fig.1). From the orbital motion we obtain for the susceptibility
\begin{equation}
\label{suc}
\chi=dM/dB=-\frac{{\mu_B^{{\rm eff}}}^2}{\hbar \Omega}
(\frac{\omega_0}{\Omega})^2 ({\sum}_++{\sum}_-)
\end{equation}
It follows from Eq.(\ref{suc}) that, for a completely filled shell,
the magnetization owing to the orbital motion leads to
diamagnetic behavior. For zero magnetic field ($k=1$) the
 system is paramagnetic and the magnetization vanishes
($\sum_- = \sum_+$). The value $k=2$ is attained at $B\approx 1.23$.
When calculating $\sum_-$ and $\sum_+$  we have to
distinguish between the cases, where the shell number $N$ of the last filled
shell is even or odd.
With all shells filled from the bottom we find (i)
 for the last filled shell number even: $ 
{\sum}_+ = {1\over 12}(N+2)[(N+2)^2+2]$, $
{\sum}_- = {1\over 6}(N+1)(N+2)(N+3)$
which implies $M=-\mu_B^{{\rm eff}}(1-\omega_L / \Omega)(N+2)/2$; and
(ii) for the last filled shell number odd:
${\sum}_+ = {1\over 2}{\sum}_- = {1\over 12}(N+1)(N+2)(N+3)$
which, in turn, implies $M=0.$
Therefore, if $\Omega_+/\Omega_-=2$, the orbital magnetization vanishes
 for the magic numbers
$4,12,24,\ldots $ while it leads to diamagnetism for the magic numbers
$2,8,18,\ldots $. A similar picture is obtained for $\Omega_+/\Omega_-=3$
which happens at $B\approx 2.01$: for each third filled shell number
(magic numbers $6,18,\ldots $) the magnetization is zero.
Since the results presented are due to shell effects, they do not depend
on the assumption $\omega _x/\omega _y=1$ which was made to facilitate the
discussion. The crucial point is the relation
$\Omega_+/\Omega_-=k=1,2,3,\ldots $ which can be obtained for a variety
of combinations of the magnetic field strength and the ratio
$\omega _x/\omega _y$ as is illustrated in Fig.2.
Whenever the appropriate combination of field strength and deformation is
chosen to yield, say, $k=2$, our findings apply.

To summarize:
the consequences of shell structure effects for the addition energy of a small
isolated quantum dot have been analyzed. At certain values of the magnetic
field strength closed shells appear in a quantum dot, also in cases where
deformation does not give rise to magic numbers at zero field strength.
Measurements of the magnetic susceptibility are expected to reflect the
properties of the single-particle spectrum and should display characteristic
patterns depending on the particle number.
At certain values of the magnetic field and electron numbers
the orbital magnetization vanishes due to shell closure
in the quantum dot.

\centerline{Figure Captions}

{\bf Fig.1} Single-particle spectra as a function of the magnetic field 
strength.
Spectra are displayed for (a) a plain isotropic and (b) deformed
two dimensional oscillator.

{\bf Fig.2} Relative magnetic field strength $B'=\omega_L/\omega _x$ as a
function of the ratio $r=\omega _x/\omega _y=1-\beta $ for fixed values of
the ratio $k=\Omega _+/\Omega _-$.

{\bf Fig.3} Magnetic susceptibility
$\chi = -\partial ^2E_{{\rm tot}}/\partial B^2$ in arbitrary units
as a function of the magnetic field strength for the isotropic oscillator
without $L^2$-term.  $E_{{\rm tot}}$ is the
sum of the single-particle energies filled from the bottom up to the electron
number $A$.

{\bf Fig.4} Blow--ups of the relevant level crossings explaining the features in
Fig.3. The left and right hand side refers to $A=8$ and $A=9$, respectively.


\begin{references}
\bibitem{Kas} M.A. Kastner, Phys.Today {\bf 46}, 24 (1993);
R.C. Ashoori, Nature (London) {\bf 379}, 413 (1996)
\bibitem{MC} P.A. Maksym and T. Chakraborty,
Phys.Rev.Lett. {\bf 65}, 108 (1990);  Phys.Rev. {\bf B45}, 1947 (1992);
M. Wagner, U.Merkt and A.V. Chaplik, Phys.Rev. {\bf B45}, 1951 (1992);
P. Hawrylak, Phys.Rev.Lett. {\bf 71}, 3347 (1993).
\bibitem{rev1} L.P. Kouwenhoven {\it et al},
{\it Mesoscopic Electron Transport, Proceed. of the NATO
ASI} L.P. Kouwenhoven, G. Sch\"{o}n and L.L. Sohn , Eds
(Series E, {\bf 345}, 105, Kluwer, Dordrecht, Netherlands, 1997)
\bibitem{NR} S.G. Nilsson and I. Ragnarsson, {\it  Shapes and Shells in
Nuclear Structure} (Cambridge, Cambridge University Press 1995)
\bibitem{HB} W.A. de Heer, Rev. Mod. Phys. {\bf 65}, 611 (1993);
M. Brack, Rev. Mod. Phys. {\bf 65}, 677 (1993).
\bibitem{e1} D.J. Lockwood {\it et al}, 
 Phys.Rev.Lett. {\bf 77}, 354 (1996).
\bibitem{e2} S.Tarucha {\it et al}, 
Phys.Rev.Lett. {\bf 77}, 3613 (1996).
\bibitem{e3} M. Fricke {\it et al}, Europ.Lett. {\bf 36}, 197 (1996).
\bibitem{Mac} M. Macucci, K. Hess and G.J. Iafrate,
Phys.Rev. {\bf B48}, 17354 (1993); J.App.Phys. {\bf 77}, 3267 (1995).
\bibitem{Mak} P.A. Maksym, Phys.Rev. {\bf B 53}, 10871 (1996).
\bibitem{PL} W.D. Heiss and R.G. Nazmitdinov,
Phys.Lett. {\bf A222}, 309 (1996).
\bibitem{Sto}  M.Stopa, Phys.Rev. {\bf B54}, 13767 (1996).
\bibitem{DN} M. Dineykhan and R.G. Nazmitdinov, Phys.Rev.{\bf B55},
13707 (1997).
\bibitem{W} A. Wojs {\it et al}, Phys.Rev. {\bf B54}, 5604 (1996).
\bibitem{ken} K.Hirose and N.S. Wingreen, cond-mat/9808193.
\bibitem{al} B.L. Altshuler, Y. Gefen and Y. Imry, Phys.Rev.Lett. {\bf 66},
88 (1991);  B.L. Altshuler {\it et al}, Phys.Rev. {\bf B47}, 10335 (1993).
\bibitem{op} F.v.Oppen, Phys.Rev. {\bf B50}, 17151 (1994).
\bibitem{ric} K.Richter, D. Ulmo and R. Jalabert,
Phys.Rep. {\bf 276}, 1 (1996).
\bibitem{W1} A. Wojs and P.Hawrylak, Phys.Rev. {\bf B53}, 10841 (1996);
M. Eto, Jpn.J.Appl.Phys. {\bf 36}, 3924 (1997); M. Koskinen,
M. Manninen and S.M. Reimann, Phys.Rev.Lett. {\bf 79}, 1389 (1997).
\bibitem{HeNa97} W.D. Heiss and R.G. Nazmitdinov,
Phys. Rev. {\bf B55}, 16310 (1997).
\bibitem{JT} H.A. Jahn and E. Teller, Proc.R.Soc.London, Sec.A {\bf 161}, 220
(1937).
\bibitem{hac} G. Hackenbroich, W.D. Heiss and H.A. Weidenm\~uller,
Phys.Rev.Lett. {\bf 79}, 127 (1997). 
\bibitem{HeNa94} W.D. Heiss and R.G. Nazmitdinov, Phys.Rev.Lett. {\bf 73},
1235 (1994); Physica {\bf D118}, 134 (1998).
\bibitem{Fock} V. Fock, Z. Phys. {\bf 47}, 446 (1928).


\end{references}
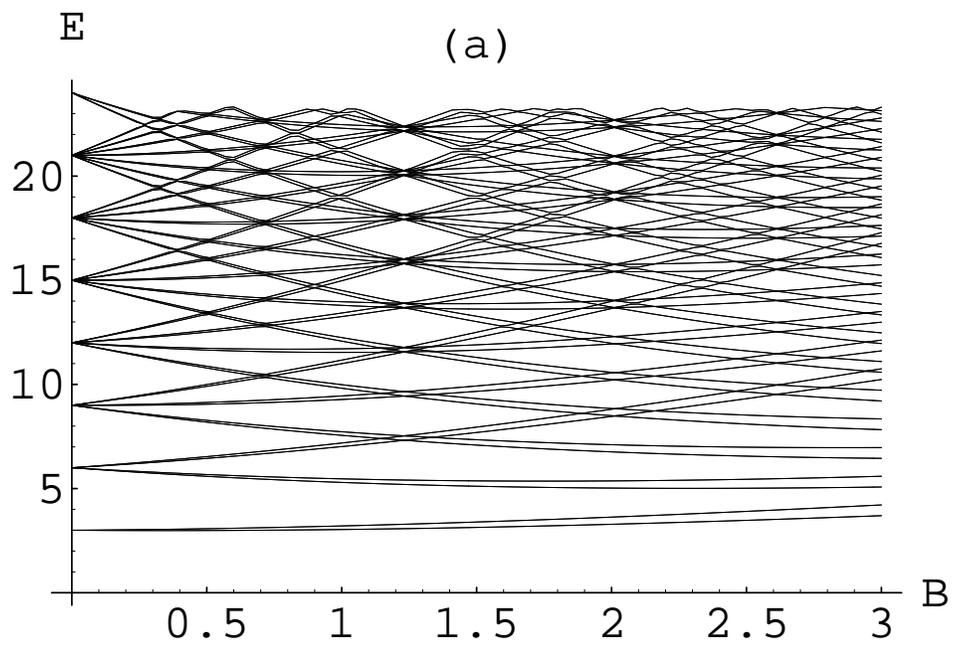
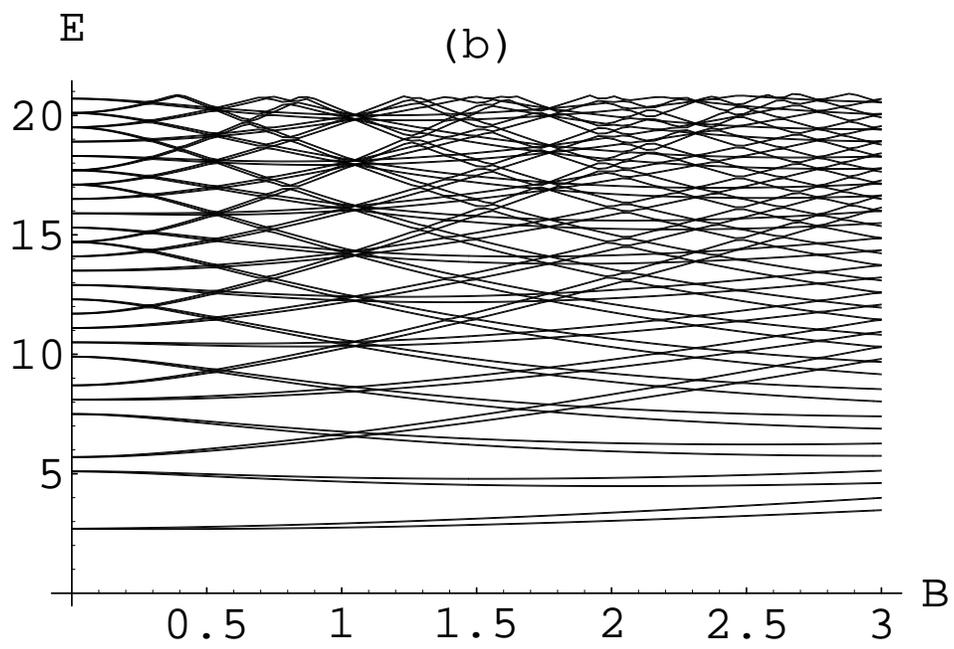
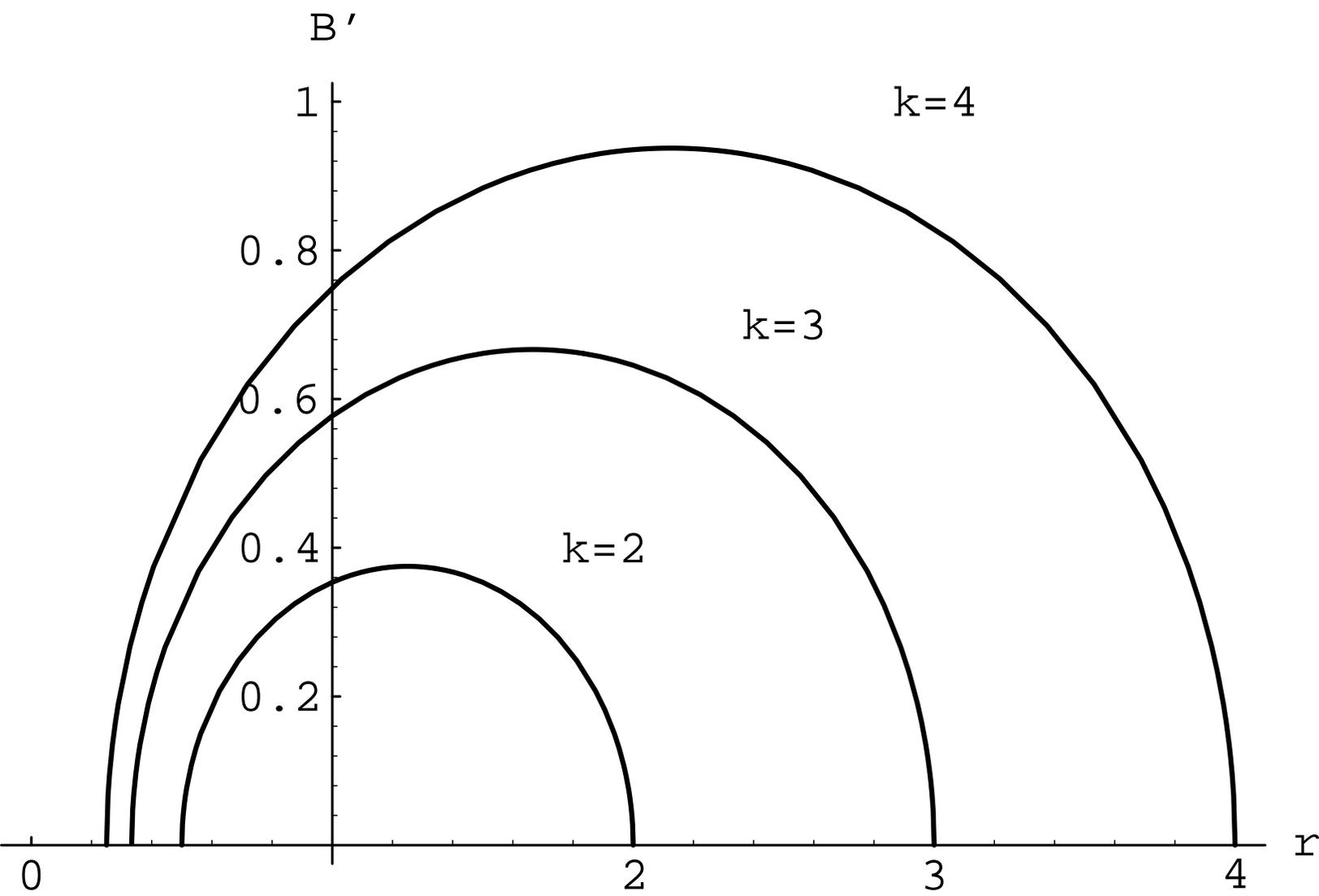
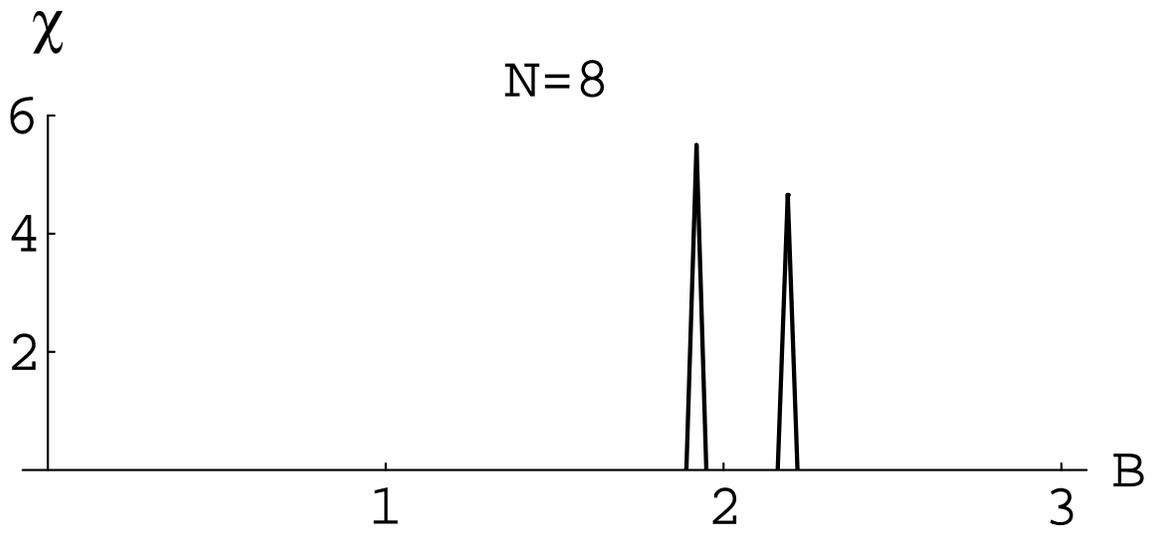
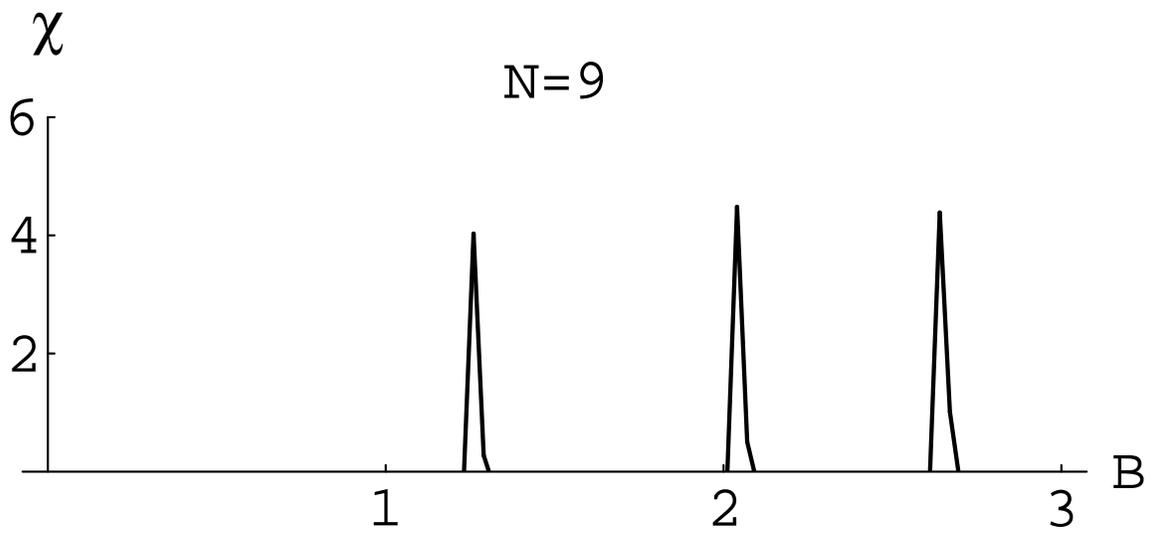
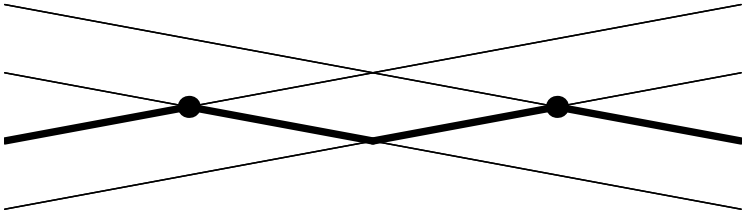
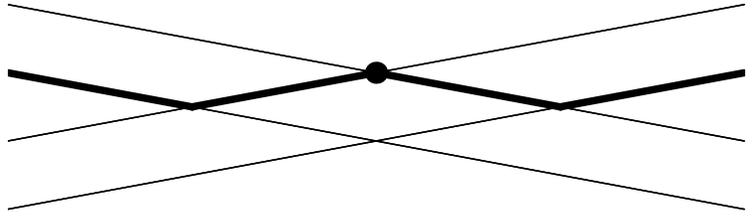
\end{document}